\documentclass[preprint2]{aastex62}
\turnoffeditone
\turnoffedittwo

\usepackage{textgreek}
\usepackage{CJKutf8}

\def\papertitle{Effects of Coronal Density and Magnetic Field Distributions on a Global Solar EUV Wave}
\def\paperkeywords{Sun: coronal mass ejections (CMEs) --- Sun: magnetic fields --- Sun: activity --- Sun: corona --- shock waves}
\def\papersubject{Sun: Extreme-ultraviolet (EUV) Wave}
\hypersetup{pdfauthor=Huidong Hu et al.,pdftitle=\papertitle,pdfkeywords=\paperkeywords,pdfsubject=\papersubject}

\def\affnssc{State Key Laboratory of Space Weather, National Space Science Center,
    Chinese Academy of Sciences, Beijing 100190, China; \href{mailto:liuxying@swl.ac.cn}{liuxying@swl.ac.cn}}
\def\affucas{University of Chinese Academy of Sciences, Beijing 100049, China}
\def\affmps{Max Planck Institute for Solar System Research, G\"{o}ttingen 37077, Germany}

\def\sdo{\emph{SDO}}
\def\soho{\emph{SOHO}}
\def\stereo{\emph{STEREO}}
\def\sta{\emph{STEREO-A}}
\def\goesr{\emph{GOES-16}}
\def\kmps{\mbox{km s$^{-1}$}}

\newcommand{\rsquo}{'}

\received{2018 November 13}
\revised{2019 April 22}
\accepted{2019 May 1}

\begin{document}
\begin{CJK}{UTF8}{gbsn}

\title{\papertitle}
\author[0000-0001-8188-9013]{Huidong Hu (胡会东)}
\affil{\affnssc}
\affil{\affmps}
\author[0000-0002-3483-5909]{Ying D. Liu (刘颍)}
\affil{\affnssc}
\affil{\affucas}
\author[0000-0001-6306-3365]{Bei Zhu (朱蓓)}
\affil{\affnssc}
\affil{\affucas}
\author[0000-0001-9921-0937]{Hardi Peter}
\affil{\affmps}
\author[0000-0003-0691-6514]{Wen He (何雯)}
\affil{\affnssc}
\affil{\affucas}
\author[0000-0001-5205-1713]{Rui Wang (王瑞)}
\affil{\affnssc}
\author[0000-0002-1509-1529]{Zhongwei Yang (杨忠炜)}
\affil{\affnssc}
\affil{\affucas}


\begin{abstract}
We investigate a global extreme-ultraviolet (EUV) wave
{associated with a coronal mass ejection (CME)-driven shock}{}
on 2017 September 10.
The EUV wave is transmitted by north- and south-polar coronal holes (CHs),
which is observed by the \emph{Solar Dynamics Observatory} (\sdo)
and \emph{Solar Terrestrial Relations Observatory A} (\sta) from opposite sides of the Sun.
We obtain key findings on how the EUV wave interacts with multiple coronal structures,
{and on its connection with the CME-driven shock:
(1) the transmitted EUV wave is still connected
with the shock that is incurvated to the Sun,
after the shock has reached the opposite side of the eruption};
(2) the south CH transmitted EUV wave is accelerated
inside an on-disk, low-density region with closed magnetic fields,
which {implies that an EUV wave
can be accelerated in both open and closed magnetic field regions};
(3) part of the primary EUV wavefront turns around a bright point (BP)
{with a bipolar magnetic structure}{}
when it approaches a dim, low-density {filament channel}{} near the BP;
(4) the primary EUV wave is diffused and apparently halted
near the boundaries of remote active regions (ARs)
{that are far from the eruption},
and no obvious {AR related}{} secondary waves are detected;
{(5) the EUV wave extends to an unprecedented scale of $\sim$360\degr{} in latitudes,
which is attributed to the polar CH transmission}.
These results provide insights into the effects of coronal density and magnetic field distributions
on the evolution of an EUV wave,
{and into the connection between the EUV wave and the associated CME-driven shock}.
\end{abstract}


\section{Introduction} \label{sec:intro}

A solar extreme-ultraviolet (EUV) wave is a large-scale intensity disturbance propagating in the corona,
which was proposed to be the coronal counterpart
\citep[e.g.,][]{ThompsonGN1999ApJ,wvm2004AA,muhrvt2010ApJ,asaiii2012ApJ}
of the ``Moreton wave''
\citep{Moreton1960AJ}.
Observations and simulations suggest that large-scale EUV waves
associated with coronal mass ejections (CMEs)
are fast-mode magnetohydrodynamic waves/shocks driven by the expanding CMEs
\citep[e.g.,][]{zhukova2004AA,cohenam2009ApJ,Patsourakosv2009ApJ,llb2011,chengzo2012ApJ,LiuON2012ApJ,ShenLS2013ApJ}.
An EUV wave will probably decouple from the driving CME after the expansion ceases in the corona,
{which is supposed to be the low-corona footprint
of an expanding fast-mode shock at the initial stage
\citep[e.g.,][]{chengzo2012ApJ,Patsourakosv2012SoPh,kzo2014ApJ}.
Observations reveal that {part of} a CME-driven shock can propagate
{in the opposite direction of the CME propagation
and reach the side of the Sun opposite to its} associated eruption
\citep[e.g.,][]{kzo2014ApJ,lhz2017apj}.
However, the connection between an EUV wave and the accompanied shock
after the latter has propagated to the opposite side is not well investigated yet.}

Coronal magnetic structures, e.g., coronal holes (CHs) and active regions (ARs),
can affect the propagation of an EUV wave.
\citet{gopalswamyvt2009ApJ} reported an EUV wave reflected from a CH.
Both reflection from and transmission through a CH have been observed by
\citet{olmedovz2012ApJ}.
\citet{ShenLS2013ApJ} reported the diffraction, refraction, and reflection of
an EUV wave interacting with remote ARs.
{Obvious secondary waves have been observed on the far side of remote ARs
that are near the eruption site in previous studies
\citep[e.g.,][]{LiZY2012ApJ,ShenLS2013ApJ}.
In this paper, we will see that the secondary wave beyond a remote AR can hardly be detected
if the AR is too far from the eruption site.
Elevation of EUV wave speed inside a dim coronal cavity of low density
on the solar limb has been reported by
\citet{LiuON2012ApJ}.
We will investigate the effect of a low-density closed magnetic field structure
on the propagation of an EUV wave, when the structure is on the solar disk.
We also find that a bright point (BP)
with a small bipolar magnetic structure
influences the wavefront of an EUV wave in combination with a filament channel.}
Large-scale EUV waves propagate to distances statistically in the range of 350--850 Mm
\citep{Patsourakosv2012SoPh}.
However, it is still unclear how large and persistent an EUV wave could be
and what contributes to its globalization and persistence.

In this paper we report a persistent, global EUV wave with a very large scale
that interacts with multiple low-density regions {with closed magnetic fields}, ARs, and CHs.
The global EUV wave was accompanied by an X8.2 flare and a violent CME {driving a shock}
on the west limb of the Sun on 2017 September 10,
which was observed by spacecraft on the opposite sides of the Sun.
The related coronal and interplanetary properties have been studied in several works
\citep[e.g.,][]{SeatonD2018ApJ,LiXD2018ApJ,YanYX2018ApJ,GoryaevSR2018ApJ,LuhmannML2018sw,LiuJD2018ApJL,ChengLW2018ApJ,LiuZZ2019APJ}.
In this paper we analyze the observations of
\emph{Solar Terrestrial Relations Observatory A} (\sta),
\emph{Solar Dynamics Observatory} (\sdo),
\emph{Geostationary Operational Environmental Satellite 16} (\goesr),
and \emph{Solar and Heliospheric Observatory} (\soho)
from nearly opposite sides of the Sun.
The time frame of this work was coincident with that of
\citet{LiuJD2018ApJL}, which is based on single-spacecraft observations.
We perform further analysis using observations from multiple spacecraft,
which covers almost 360\degr{} of the Sun.
This paper focuses on
{the connection with the accompanied CME-driven shock at a later stage;}
the alterations of the speed and wavefront
caused by low-density regions, {a BP,} ARs, and CHs;
{and the long persistence of the EUV wave.
The global properties of the EUV wave and
its connection with the CME-driven shock observed from two sides of the Sun
are investigated in Section \ref{sec:global}.
The interactions with coronal structures
and the global scale of the EUV wave are studied in Section \ref{sec:interactions}.
We conclude and discuss the results in Section \ref{sec:concl}.}
Our results {improve the knowledge of the connection with the CME-driven shock and}
highlight the effects of coronal density and magnetic field distributions
on the propagation and globalization of an EUV wave.

\begin{figure*}
\epsscale{0.9}
\plotone{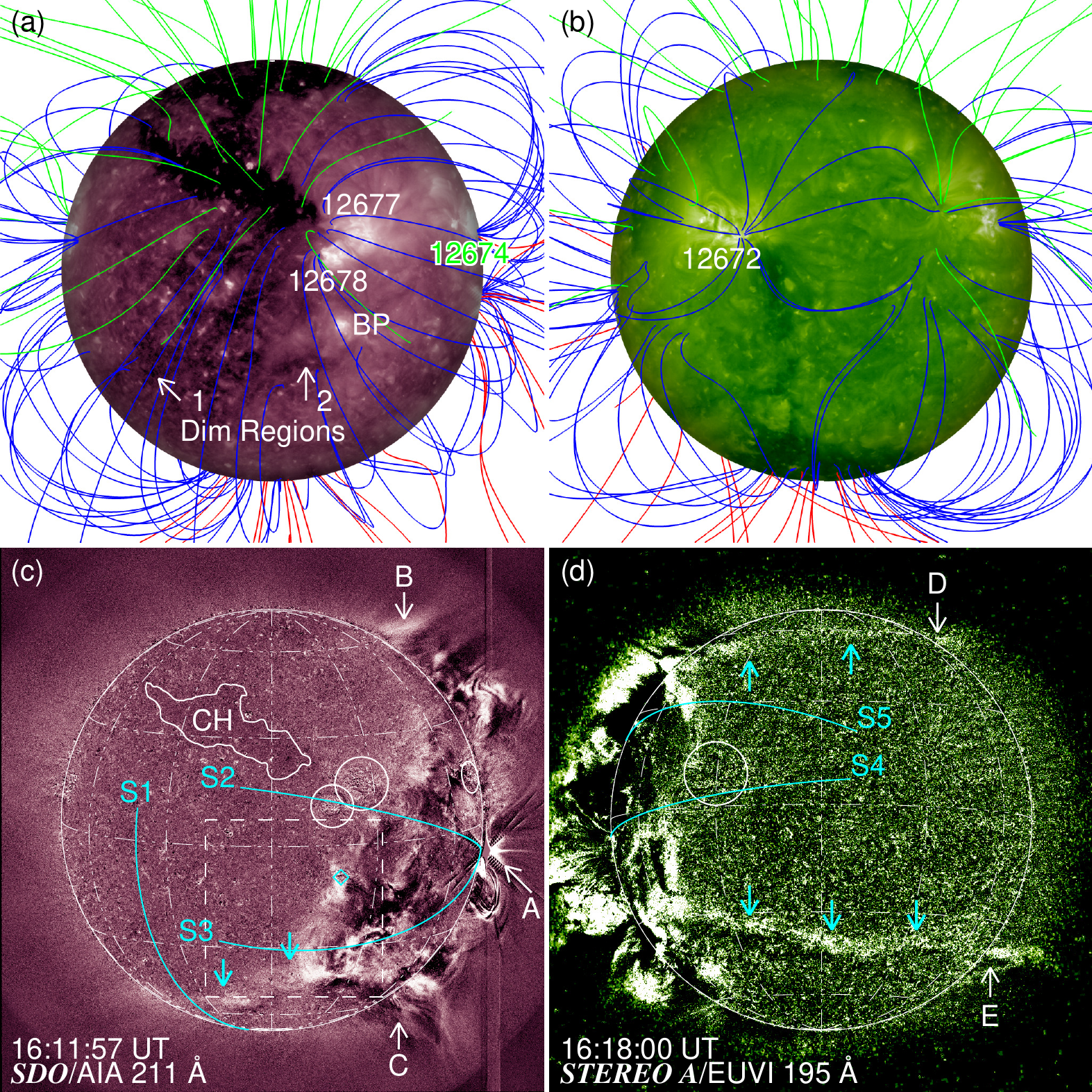}
\caption{\label{fig:euv_pfss}EUV observations and coronal magnetic field distributions.
(a-b) Images of \sdo/AIA 211 \AA{} and \sta/EUVI 195 \AA{}
with coronal magnetic field lines obtained from a PFSS model.
The blue lines represent the closed magnetic field lines,
and the green (red) lines denote the positive (negative) open field lines.
Arrows ``1'' and ``2'' indicate two dim regions, and ``BP'' marks a bright point.
(c) Running-difference image of \sdo/AIA 211 \AA.
Arrow ``A'' points to the flare.
Arrows ``B'' and ``C'' indicate the approximate sites where the EUV wave interacts with the polar CHs.
The closed region denoted with ``CH'' represents a CH in the north midlatitude area.
The diamond marks the bright point.
The dashed square specifies an area for Figure \ref{fig:turn}.
(d) Running-difference image of \sta/EUVI 195 \AA,
{of which the FOV is slightly changed
to be comparable to that of panel (c).}
Arrows ``D'' and ``E'' denote the polar CH transmitted waves on the limb.
{Cyan} arrows in panels (c-d) indicate the on-disk wavefronts of the transmitted waves.
``S1'' is a great-circle slit along the E40\degr{} meridian,
and ``S2-5'' are great-circle slits starting from the flare site.
The circles/ellipse in panels (c-d) represent ARs.
{The heliographic grids are spaced by 30\degr.}
{An animation of panels (c-d) is available,
which begins at 15:40 UT and ends at 17:26 UT.
In the left panel of the animation,
running-difference images of \sdo/AIA 211, 193, and 171 \AA{}
are synthesized to red-green-blue (RGB) color images.
The right panel of the animation shows
the running-difference images of \sta/EUVI with the original FOV.
The lines, arrows, and annotations are removed in the animation.}\\
(An animation of this figure is available.)}
\end{figure*}

\section{Global Properties and Associated Shock} \label{sec:global}
The global EUV wave is associated with an energetic CME {driving a shock}
and an X8.2 flare from AR 12673 (S08\degr W88\degr)
that peaked at 16:06 UT on 2017 September 10.
The Atmospheric Imaging Assembly (AIA)
on board \sdo{}
\citep{lta_aia2012} and
the Extreme Ultraviolet Imager (EUVI) on board \stereo{}
\citep{kkd_stereo2008} are advantageous
when observing the Sun in EUV bands from multiple viewpoints
\citep[e.g.,][]{Patsourakosv2009ApJ,llb2011,llk2014NatCo,olmedovz2012ApJ,HuLW2017ApJ,ZhuLK2018APJ}.
Figure \ref{fig:euv_pfss} shows the EUV observations of \sdo/AIA 211 \AA{} near the Earth
and \sta/EUVI{ (EUVI-A)} 195 \AA{} from $\sim$128\degr{} east of the Earth.
The dark regions in the AIA 211 \AA{} (Figure \ref{fig:euv_pfss}a)
and EUVI-A 195 \AA{} (Figure \ref{fig:euv_pfss}b) images
illustrate three CHs in the north- and south-polar regions as well as a northern midlatitude area.
The open magnetic field lines (green and red lines in Figure \ref{fig:euv_pfss}a-b)
derived from a potential-field source-surface (PFSS) model also demonstrate the CHs.
``BP'' marks a BP that has a bipolar magnetic structure.
Arrows ``1'' and ``2'' in Figure \ref{fig:euv_pfss}(a) denote two dim regions
where the coronal density and {possibly also the temperature are} relatively low.
The running-difference (RD) image of AIA 211 \AA{} (Figure \ref{fig:euv_pfss}c) shows that
the EUV wave was propagating on the western solar disk away from the flare site (indicated by arrow ``A'') at 16:11 UT.
The EUV wave was interacting with the polar CHs
near 60\degr{} north and south latitudes on the west limb,
as indicated by arrows ``B'' and ``C''.
RD images of AIA 211, 193, and 171 \AA{} are synthesized to red-green-blue color images
and then join the RD images of EUVI-A 195 \AA{} (similar to Figure \ref{fig:euv_pfss}d)
to make {an animation for Figure \ref{fig:euv_pfss}(c-d)}.
The animation illustrates that the EUV wave is reflected and also transmitted by both the north- and south-polar CHs,
which is also reported by
\citet{LiuJD2018ApJL}.
Two antipoleward transmitted waves emanate along the boundaries of the two polar CHs,
which are indicated by the {cyan} arrows in Figure \ref{fig:euv_pfss}(c-d).
Furthermore, the north transmitted wave is transmitted once again by the midlatitude CH
(marked with ``CH'' in Figure \ref{fig:euv_pfss}c)
and produces a secondary wave along the southern boundary of the CH,
which is visible in the animation of Figure \ref{fig:euv_pfss}.
The wavefronts of the transmitted waves on the limb (denoted with arrows ``D'' and ``E'')
are forwardly inclined,
which are also observed on the east limb by AIA in the animation of Figure \ref{fig:euv_pfss}.
Note that there is no driving CME structure following the \emph{transmitted} EUV wave,
which is different from the explanation with the downward expansion of a CME structure in
\citet{LiuON2012ApJ}.
This suggests that the wave speed increases with the altitude in the low corona,
which is consistent with the model of \citet{Uchida1968SoPh}.
The CME structure and the \emph{primary} EUV wavefront are obvious
in the EUVI-A images from 15:55 to 15:58 UT in the animation of Figure \ref{fig:euv_pfss},
which agrees with a primary EUV wave being a fast-mode wave/shock driven by the associated CME
\citep[e.g.,][]{Patsourakosv2009ApJ,llb2011,chengzo2012ApJ,DownsRH2012ApJ}.
As a shock signature
\citep[e.g.,][]{WarmuthVM2004AA,llb2009,nst2013ApJ,hlw2016ApJ,YangLL2018ApJ},
a type II radio burst associated with our case is also detected by ground-based radio spectrometers
\citep{GopalswamyYM2018ApJL}.

\begin{figure*}
\epsscale{.85}
\plotone{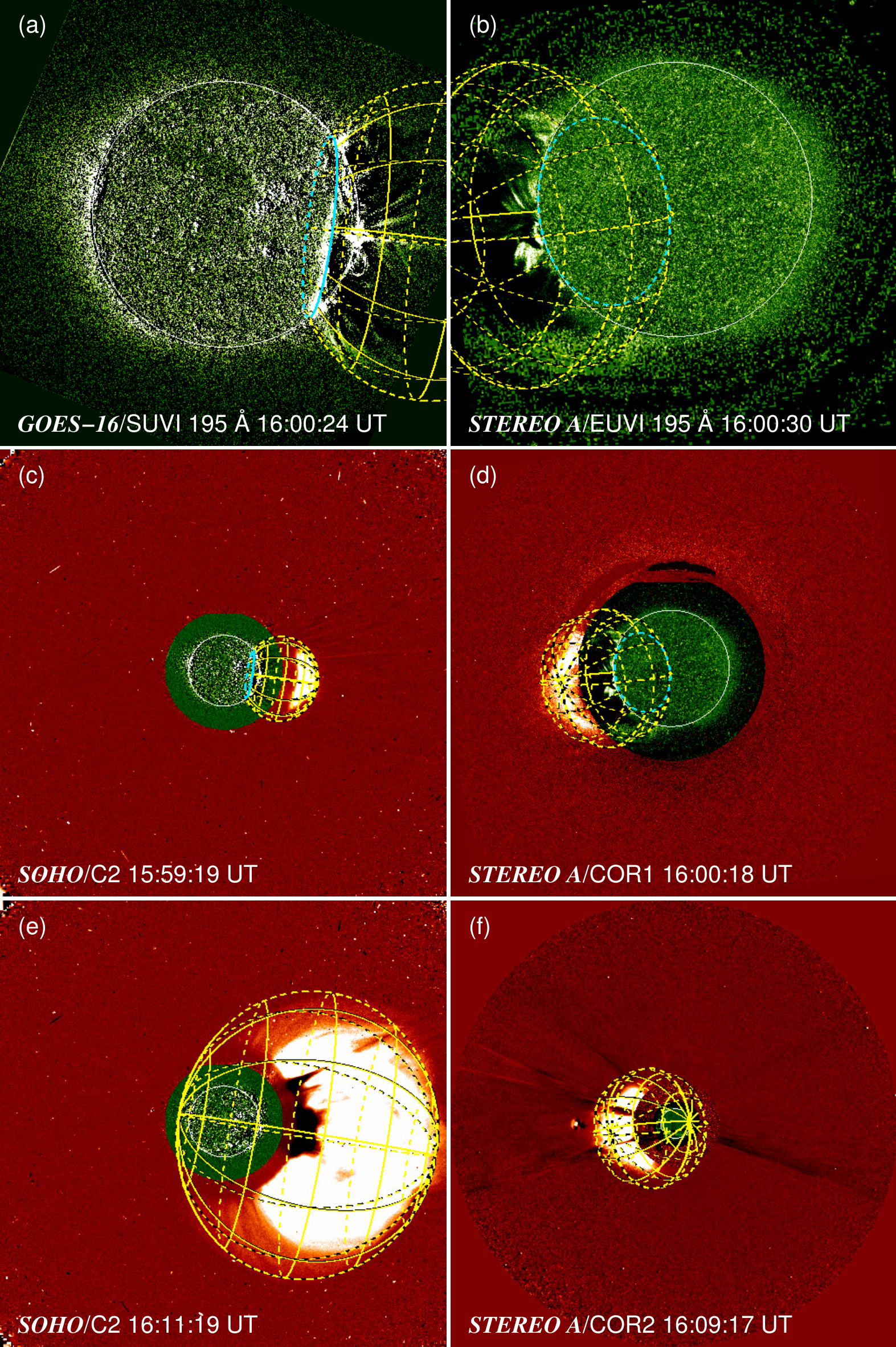}
\caption{\label{fig:shock}{Running-difference images from \goesr, \soho, and \sta{}
superimposed with a three-dimensional modeled shock.
The overlaid frame represents the constructed shock
with the ellipsoid model developed by \citet{kzo2014ApJ}.
{Panels (a-d) show the same modeled result for 16:00 UT.
The cyan ellipse indicates the intersection of the model ellipsoid and the Sun surface.}
The dashed curves demonstrate the part of the ellipsoid surface
{and the intersection that are} on the far side of the image plane.}}
\end{figure*}

{A large CME-driven shock is observed by coronagraphs on board \sta{} and \soho,
which propagates to the opposite side of the eruption.
\citet{LiuZZ2019APJ} present a study of the geometry and kinematics
of the CME-driven shock in relation to its heliospheric impacts
combining remote-sensing and in situ observations.
Here we focus on the connection with the EUV wave
and the morphology of the CME-driven shock
on the opposite side of the eruption,
which have not been fully understood yet.
A three-dimensional (3D) modeled structure of the shock
based on imaging observations from \sta{}, \soho, and \goesr{}
is illustrated in Figure \ref{fig:shock}
(also see \citealt{LiuZZ2019APJ}).
The 3D model is developed by \citet{kzo2014ApJ}
using an ellipsoid structure to represent a CME-driven shock,
which can determine the 3D outermost structure of the shock
by fitting multiple-viewpoint imaging observations
\citep[e.g.,][]{KwonV2017ApJ,ZhuLK2018APJ,LiuZZ2019APJ}.
The ellipsoid model has seven free parameters:
the height, longitude, and latitude of the center of the ellipsoid;
the lengths of the three semiprincipal axes; and the rotation angle of the ellipsoid.
In our fitting, the two semiprincipal axes perpendicular to the radial direction are set to equal,
which reduces two free parameters (one semiprincipal axis and the rotation angle).
We fit the parameters by visually matching the ellipsoid structure
and the shock feature in the RD images from two viewpoints as shown in Figure \ref{fig:shock}.
Images of \goesr{} Solar Ultraviolet Imager (SUVI)
with a larger field of view (FOV) instead of those of \sdo/AIA are used in the fitting.
At 16:00 UT, the lateral boundary of the ellipsoid is determined by both the EUV
(panels a-b of Figure \ref{fig:shock})
and coronagraph images (panels c-d),
and the radial boundary is restricted by the coronagraph images.
As shown in Figure \ref{fig:shock}(a),
the intersection (the solid {cyan} curve on the disk) of the ellipsoid and the Sun surface
is consistent with the primary EUV wavefront,
which agrees with a primary EUV wave being the footprint of a CME-driven shock at the initial stage
\citep[e.g.,][]{LiZY2012ApJ,kzo2014ApJ}.
{In Figure \ref{fig:shock}(b),
the dashed cyan ellipse illustrates that the intersection is on the side opposite to \sta{},
which is consistent with that no on-disk EUV wave was detected by EUVI-A around 16:00 UT.
\citet{VeronigPD2018ApJ} suggest that the EUV wave observed by SUVI
is connected with the shock detected by C2 and detaches from the CME structure in the initial stage.
The shock is caused by the impulsive expansion of the CME structure \citep{VeronigPD2018ApJ,LiuZZ2019APJ}.
Therefore, it is still appropriate to call the shock a ``CME-driven'' shock
despite the early detachment of the shock from the CME structure.
In this paper we fit only the CME-driven shock using the 3D ellipsoid model.
Investigation of the CME geometry can be seen in, e.g.,
\citet{GopalswamyYM2018ApJL}, \citet{VeronigPD2018ApJ}, and \citet{LiuZZ2019APJ}.}
Figure \ref{fig:shock}(e-f) present the modeled shock structure at $\sim$16:11 UT,
and show that the shock has almost propagated backward to the opposite side of the eruption.
However, on that time the primary EUV wave was still in the western hemisphere of the Sun
(see Figure \ref{fig:euv_pfss}c).}
The primary EUV wave lasted about 40 minutes from $\sim$15:50 UT to $\sim$16:30 UT,
and the transmitted waves started around 16:06 UT and faded out after 17:00 UT.
It is surprising that the EUV waves can persist for $\sim$50 minutes
even after {the CME-driven shock has propagated to the opposite side of the eruption.}

\begin{figure*}
\epsscale{1}
\plotone{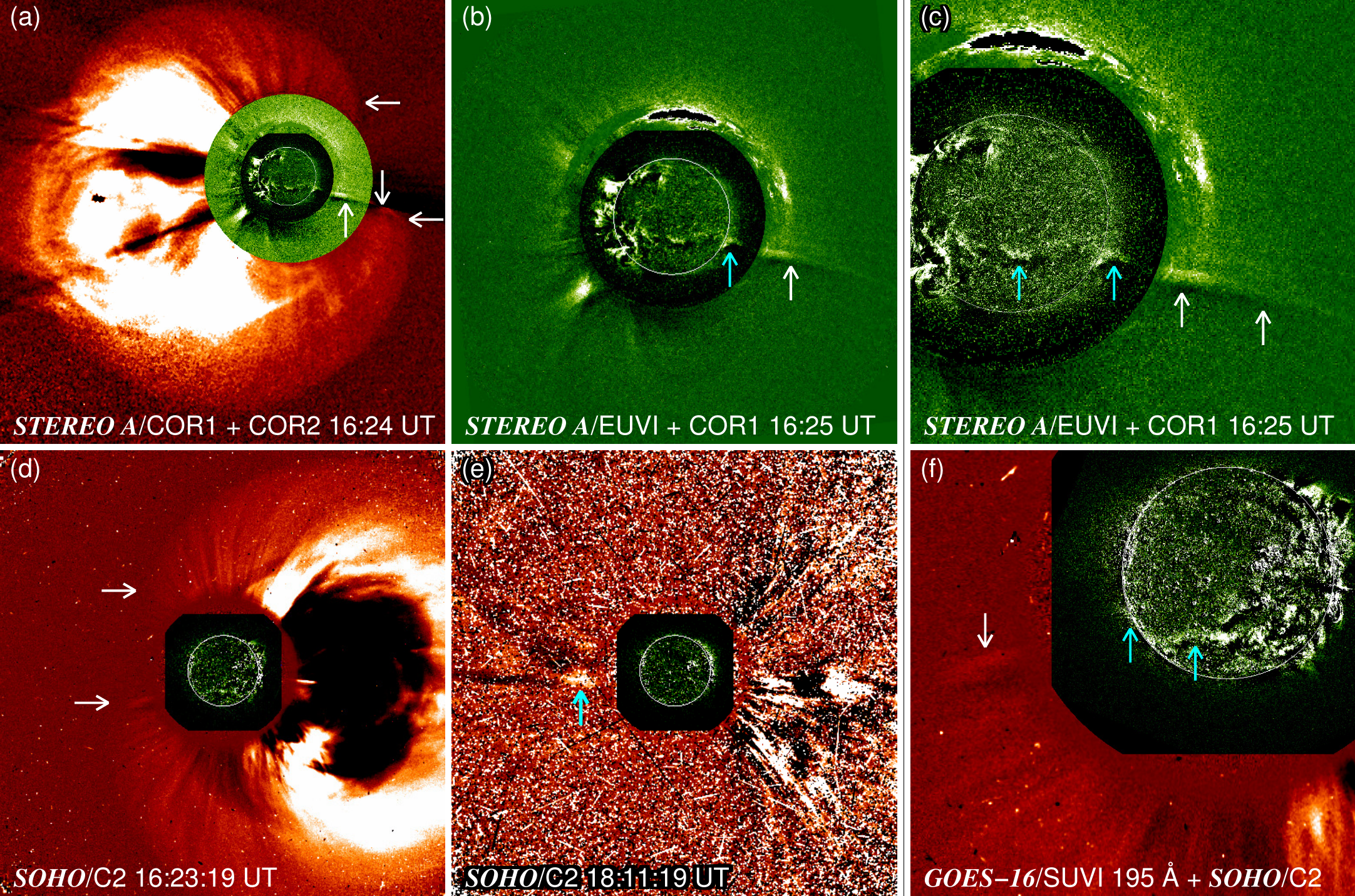}
\caption{\label{fig:backward}Running-difference images
from \soho{} and \sta{} showing the connection between the EUV wave
{and the shock on the opposite side of the eruption.
(a) Enlarged view of a \sta{}/COR2 image at 16:24 UT
superimposed with a COR1 image at 16:25 UT.}
Arrows in panel ({a}) indicate the ``incurvated'' shock
on the opposite side of the eruption {observed by COR1 and COR2}.
({b}) COR1 image {blended} with an EUVI image.
{(c) Enlarged view of panel (b).
Arrows in panels (b-c) mark the transmitted EUV wave observed by EUVI
and the southern lateral shock detected by COR1.}
({d-e}) Images of \soho{}/C2
overlaid with \goesr/SUVI images.
{Arrows in panel (d) indicate the shock on the opposite side of the eruption.}
A collision between two streamers
is denoted by the arrow in panel ({e}),
which is pronounced in {the animation of panels (d-e).
The animation begins at 15:35 UT and ends at 18:35 UT.
The arrows in panels (d-e) are removed in the animation.}
(f) Enlarged view of panel (d).
Arrows in panel (f) mark the shock observed by C2
and the transmitted EUV wave detected by SUVI.
Note that the transmitted EUV wave on the limb indicated by the middle arrow in panel (f)
is more obvious in the animation of Figure \ref{fig:euv_pfss}.\\
(An animation of this figure is available.)}
\end{figure*}

{Figure \ref{fig:backward} depicts
the connection with the EUV wave and the morphology of the backward propagating shock.
As indicated by the two {horizontal arrows in panel (a)},
the RD image of \sta{}/COR2 clearly displays that
the northern and southern parts of the shock on the opposite side of the eruption
are ``incurvated'' to the Sun and form a ``funnel-like'' structure.
The incurvated southern part of the shock observed by COR2 around 16:24 UT
is cospatial with the lateral shock observed by COR1
{as marked by the two vertical arrows in panel (a)}.
The RD COR1 image overlaid with a RD EUVI image in panel ({b})
({which is enlarged in panel c})
obviously shows that the southern lateral shock in the COR1 image
also spatially corresponds to the aforementioned south-polar CH transmitted EUV wave.
It is indicated that the incurvated shock observed by the coronagraphs on board \sta{}
is still connected with the transmitted EUV wave on the opposite side of the eruption.
The incurvation can also be seen in the RD \soho{}/C2 image in panel ({d}),
although the shock feature is relatively faint.
{The spatial correspondence between
the transmitted EUV wave and the lateral shock is not definitely observed
as shown in Figure \ref{fig:backward}(f).
However, the transmitted EUV wave on the limb is approximately radially aligned with the shock,
which also suggests that they are connected.}
A corresponding northern lateral shock is not visible in the RD COR1 image,
which is probably because it is relatively weak.
However, the animation of Figure \ref{fig:backward} of RD \soho{/C2} images
shows a lateral collision between two coronal structures
on the opposite side of the eruption around 18:11 UT,
which is also marked by the arrow in Figure \ref{fig:backward}({e}).
The two colliding structures are probably streamers
pushed by the southern and northern lateral parts of the shock.
(Streamers on the opposite side of the eruption
are indeed observed in C2 images that are not presented here.)
The joint observation of \sta/EUVI, COR1, and COR2 reveals that
the CME-driven shock is incurvated to the Sun and
still connected with the transmitted EUV wave
on the opposite side of the eruption.}

\begin{figure}
\epsscale{0.9}
\plotone{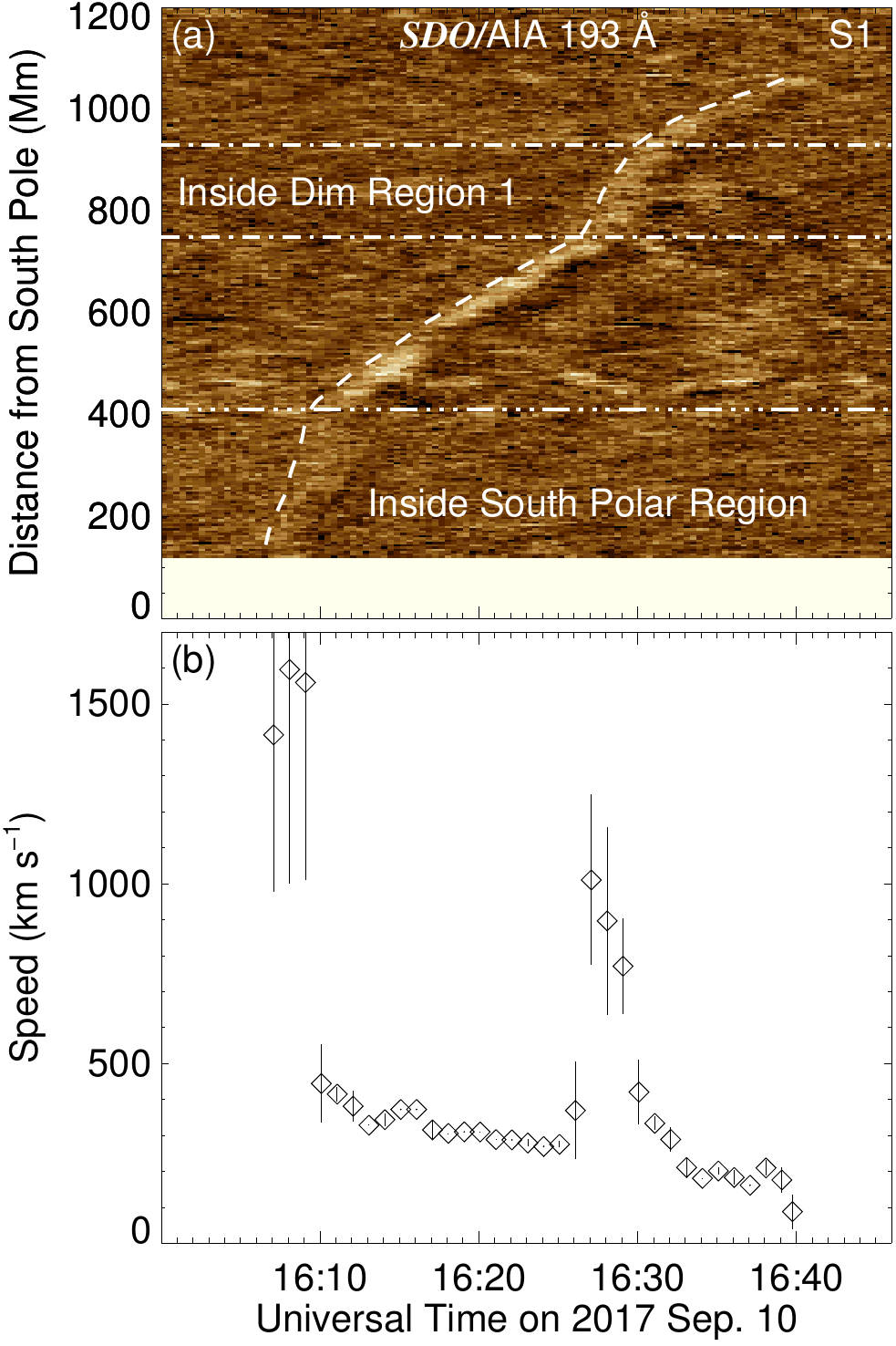}
\caption{\label{fig:mer_jmap}(a) Distance-time profile of the transmitted EUV wave
by stacking \sdo/AIA 193 \AA{} running-difference images along ``S1''.
The dashed curve indicates the wavefront,
{along which the distance is extracted}.
The top two lines mark the boundaries of {Dim Region 1} on the path,
and the third line denotes the boundary of the south-polar CH.
(b) The speed of the wave derived from the distance-time profile
{using a numerical differentiation with three-point Lagrangian interpolation,
which is then binned to reduce the scatter.}}
\end{figure}

\begin{figure}
\epsscale{0.9}
\plotone{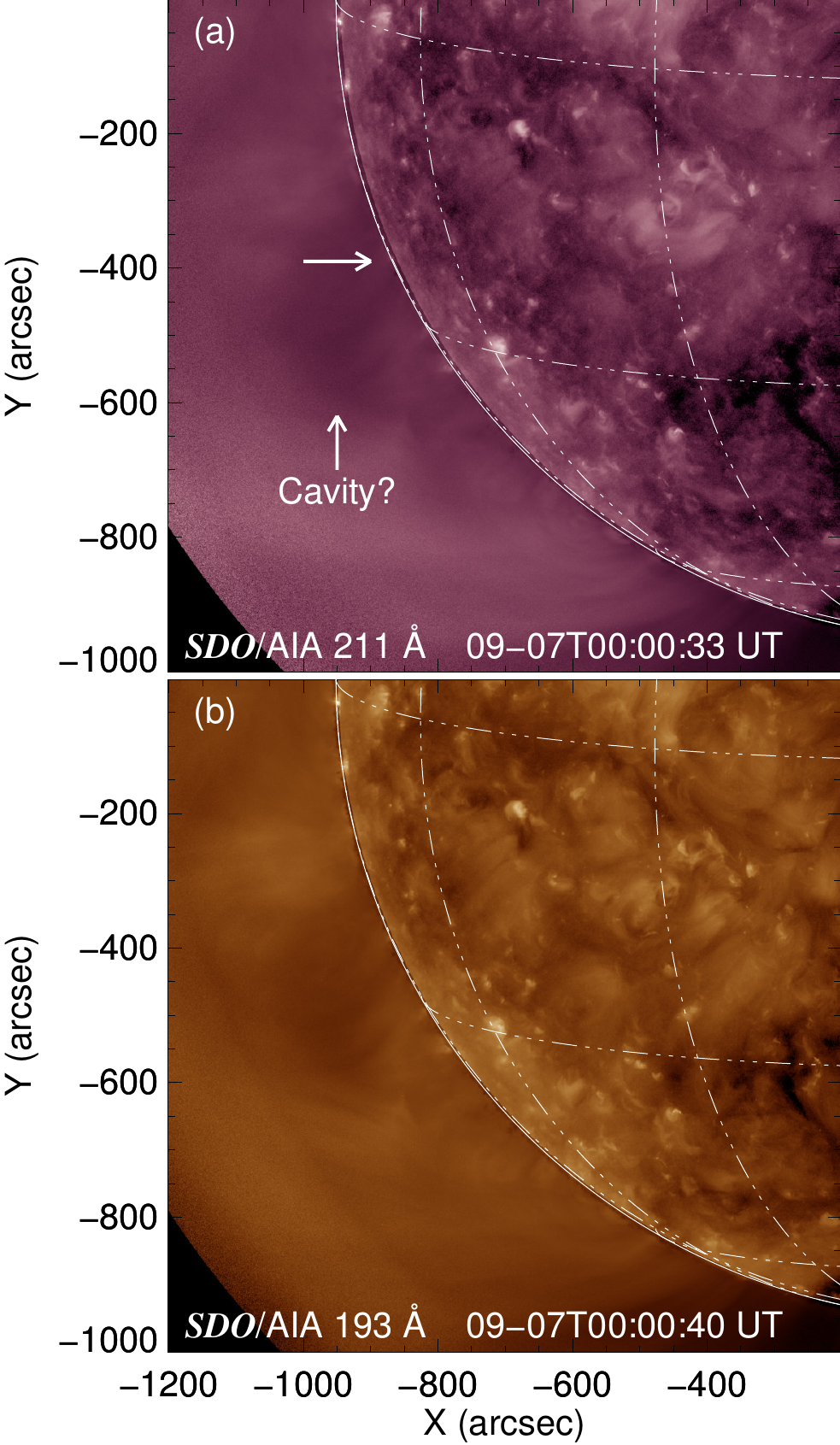}
\caption{\label{fig:cavity}{\sdo/AIA 211 and 193 \AA{} images at 00:00 UT on 2017 September 7
when Dim Region 1 was traversing the east limb.
The contrast of the off-limb corona is enhanced by using the SolarSoft routine \emph{aia\_rfilter}
(see \url{http://aia.cfa.harvard.edu/rfilter.shtml} for details).
The vertical arrow indicates a plausible coronal cavity, and
the horizontal arrow denotes the approximate location of Dim Region 1.
The heliographic grids are spaced by 30\degr.}}
\end{figure}

\section{Interactions with Dim Regions and Active Regions} \label{sec:interactions}
Five great-circle slits ``S1-5'' ({cyan} lines in Figure \ref{fig:euv_pfss}c-d)
of RD EUV images are extracted and then stacked in time sequences
to investigate the propagation of the EUV wave.
Figure \ref{fig:mer_jmap} shows the distance-time profile and the derived speed
of the south-polar CH transmitted wave along ``S1''.
{We employ a three-point Lagrangian interpolation
to obtain the speed of the EUV wave
and use the approximate values,
which is adequate to analyze the general speed variation in this paper,
although the interpolation may not be sufficiently accurate as suggested by
\citet{ByrneLG2013AA}.}
The transmitted wave was traveling inside the south-polar region
with a high local characteristic wave speed before $\sim$16:10 UT.
The wave speed dropped from $\sim$1500 \kmps{} to below 500 \kmps{}
after the transmitted wave left the south-polar region.
However, the wave was unexpectedly accelerated to nearly 1000 \kmps{} around 16:27 UT.
The acceleration is associated with an AIA 211 \AA{} dim region
({Dim Region 1,} indicated by arrow ``1'' in Figure \ref{fig:euv_pfss}a)
{that is also dimmer than its neighboring areas in the 171 and 193 \AA{} bands.
The lower EUV intensity suggests that the density and possibly the temperature
are lower in the dim region.
A low density can cause a high characteristic wave speed
by increasing the Alfv\'{e}n speed.
However, the possible low temperature can hardly decrease the speed
in the low corona with plasma beta $\beta < 1$.
The acceleration around 16:27 UT as shown in Figure \ref{fig:mer_jmap}
is consistent with the high characteristic speed inside the dim region.
The acceleration inside the dim region}
is not similar to that caused by a bright AR
\citep[e.g.,][]{LiZY2012ApJ}.
As shown by the PFSS modeling results in Figure \ref{fig:euv_pfss}(a),
the dim region is with closed magnetic fields and is also different from a CH.
{No filament associated with Dim Region 1
is observed in GONG H\textalpha{} images.
As displayed in Figure \ref{fig:cavity},
we enhance the contrast of the off-limb corona in AIA 211 and 193 \AA{} images
at the beginning of September 7 (when Dim Region 1 was traversing the east limb).
With no obvious prominence signature, a plausible cavity (indicated by the vertical arrow)
was above the corresponding site of the dim region (indicated by the horizontal arrow).
This implies that Dim Region 1 with closed magnetic fields
is the on-disk signature of a cavity in the corona,
assuming that the coronal structure did not change much in the past three days.}
\citet{LiuON2012ApJ} note that an off-limb coronal cavity could elevate the EUV wave speed.
{Acceleration associated with an on-disk dim region with closed magnetic fields
similar to our case has not been reported previously.
In our case, the acceleration may be caused by the on-disk low-density cavity with closed magnetic fields,
although the cavity is not a typical one embedding a filament
\citep[e.g.,][]{LiuON2012ApJ,ForlandGD2013SoPh}.}

\begin{figure*}
\epsscale{1.1}
\plotone{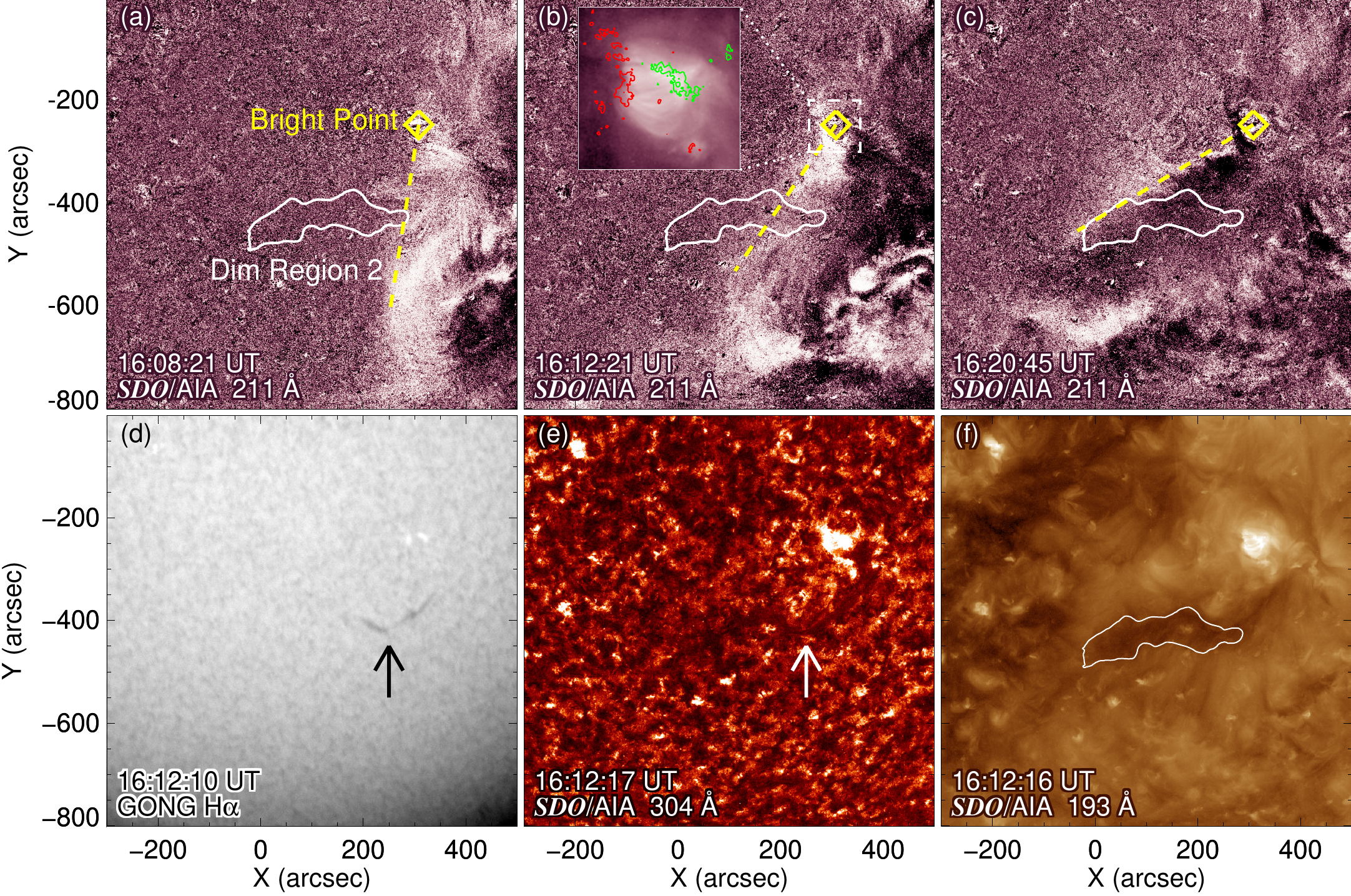}
\caption{\label{fig:turn}Observations of the square region specified in Figure \ref{fig:euv_pfss}(c).
(a-c) \sdo/AIA 211 \AA{} running-difference images.
The diamond indicates the bright point.
The dashed line denotes the wavefront,
{which is determined visually from the enhanced intensity of the running-difference images.
The inset overlaid in panel (b) is an enlarged view of the bright point
superimposed with contours of positive (green) and negative (red) 80 gauss magnetic fields.
(d-f) GONG H\textalpha{} and \sdo/AIA 304 and 193 \AA{} images.
The arrow indicates a filament.
The white closed curve delineates Dim Region 2 associated with the filament channel.
The dimming in the bottom right corner of panel (d) is due to the limb darkening
\citep{White1962ApJS}.}}
\end{figure*}

Figure \ref{fig:turn} presents {the EUV and H\textalpha{} imaging observations} of an area
specified by the dashed square in Figure \ref{fig:euv_pfss}(c).
{The diamond represents the BP
that has a bipolar magnetic structure resembling a small AR
as displayed in the inset overlaid in panel (b).
In the inset the green and red closed curves
indicate the positive and negative polarities of the magnetic field,
which are linked by bright loops.
The BP emerged around 23:00 UT on September 8
and started to fade on September 13.
The white closed curve in Figure \ref{fig:turn} delineates another dim region
({Dim Region 2,} denoted with arrow ``2'' in Figure \ref{fig:euv_pfss}a).
A filament associated with Dim Region 2 is observed
in the GONG H\textalpha{} and AIA 304 \AA{} images
(indicated by the arrows in Figure \ref{fig:turn}d-e).
The AIA 193 \AA{} image in panel (f) illustrates that
Dim Region 2 corresponds to the filament channel.
The low EUV intensity indicates that Dim Region 2 is also
of low density and has a high characteristic wave speed like Dim Region 1.}
The dashed line in Figure \ref{fig:turn}(a-c) represents the approximate primary EUV wavefront,
{which is visually identified from the enhanced intensity in the AIA 211 \AA{} RD images.
The wavefront} turns $\sim$50\degr{} around the BP from 16:08 to 16:20 UT.
The EUV wave inside the low-density {filament channel} propagates faster than outside,
whereas the wave interacting with the BP (a bipolar magnetic structure) is halted.
These cause different speeds for different parts of the wavefront
and eventually make the wavefront direction turn around the BP.

\begin{figure*}
\epsscale{0.9}
\plotone{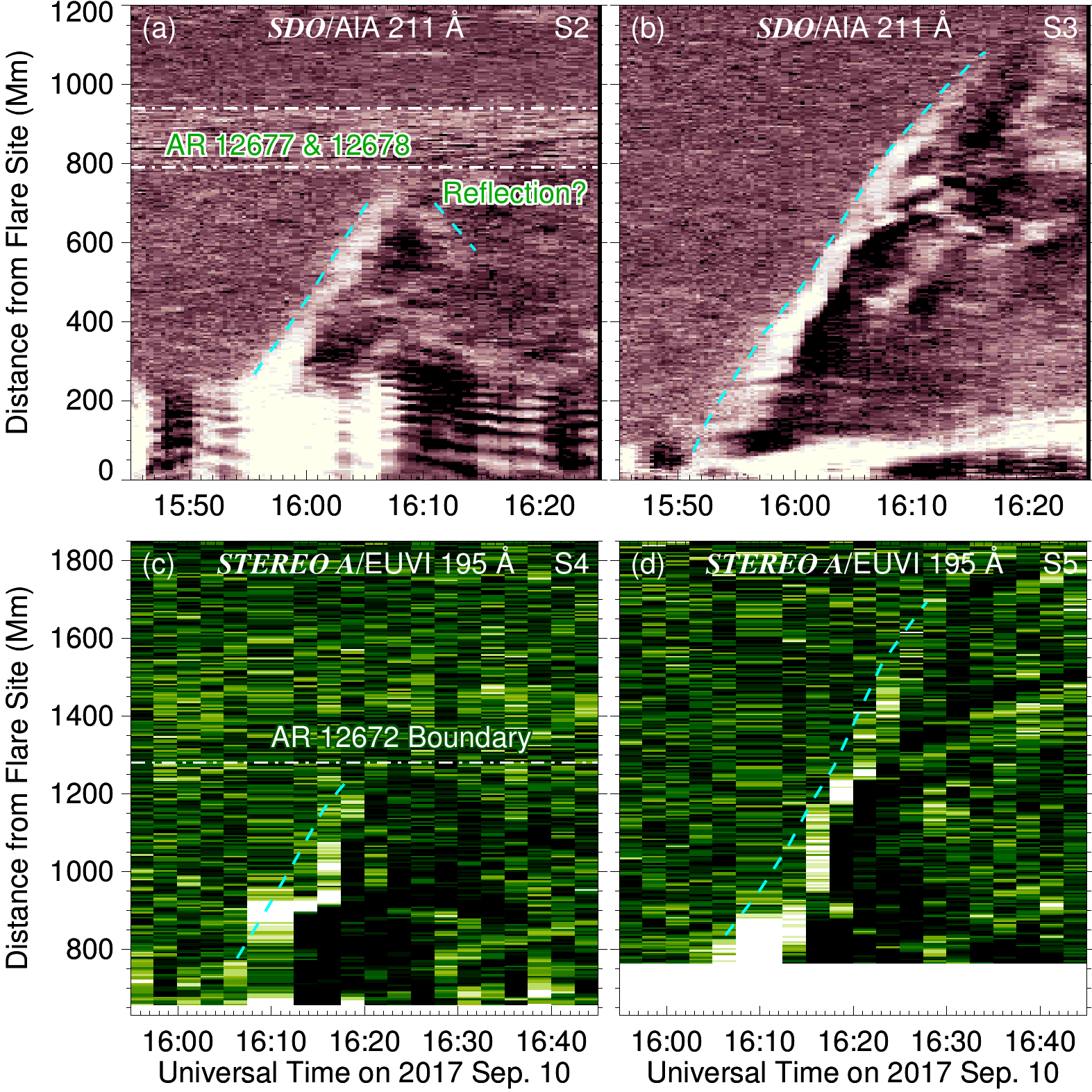}
\caption{\label{fig:ar_jmap}Distance-time profiles of the primary EUV wave
by stacking \sdo/AIA 211 \AA{} and \sta/EUVI 195 \AA{} running-difference images along ``S2-5''.
The dashed curves indicate the wavefront, and the horizontal lines mark the AR boundaries.}
\end{figure*}

Figure \ref{fig:ar_jmap} displays the distance-time profiles of the primary EUV wave along slits ``S2-5''
from the observations of AIA 211 \AA{} (panels a-b) and EUVI-A 195 \AA{} (panels c-d).
As shown in panel (a),
the primary EUV wave is diffused and possibly reflected
near the joint boundary of adjacent ARs 12677 and 12678 at $\sim$800 Mm from the flare site.
In contrast, the primary wave travels to 1100 Mm along ``S3'' without ARs on its path.
The primary wave vanishes near the boundary of AR 12672 at $\sim$1300 Mm along ``S4'',
but propagates as far as 1700 Mm along ``S5'' without ARs on its path.
The average speeds along ``S2-5'' are 600--700 \kmps.
No obvious transmitted secondary wave is observed on the far side of the ARs,
which is different from the cases in
\citet{LiZY2012ApJ} and \citet{ShenLS2013ApJ}
with remote ARs within 400 Mm from the eruptions.
However, the animation of Figure \ref{fig:euv_pfss} shows that
the primary EUV wave continues propagating after it passes AR 12674.
AR 12674 is close to the flare site with a distance of $\sim$300 Mm,
where the expanding CME is expected to be still driving the EUV wave.
This may be the reason that the EUV wave travels beyond AR 12674.

\begin{figure*}
\epsscale{0.8}
\plotone{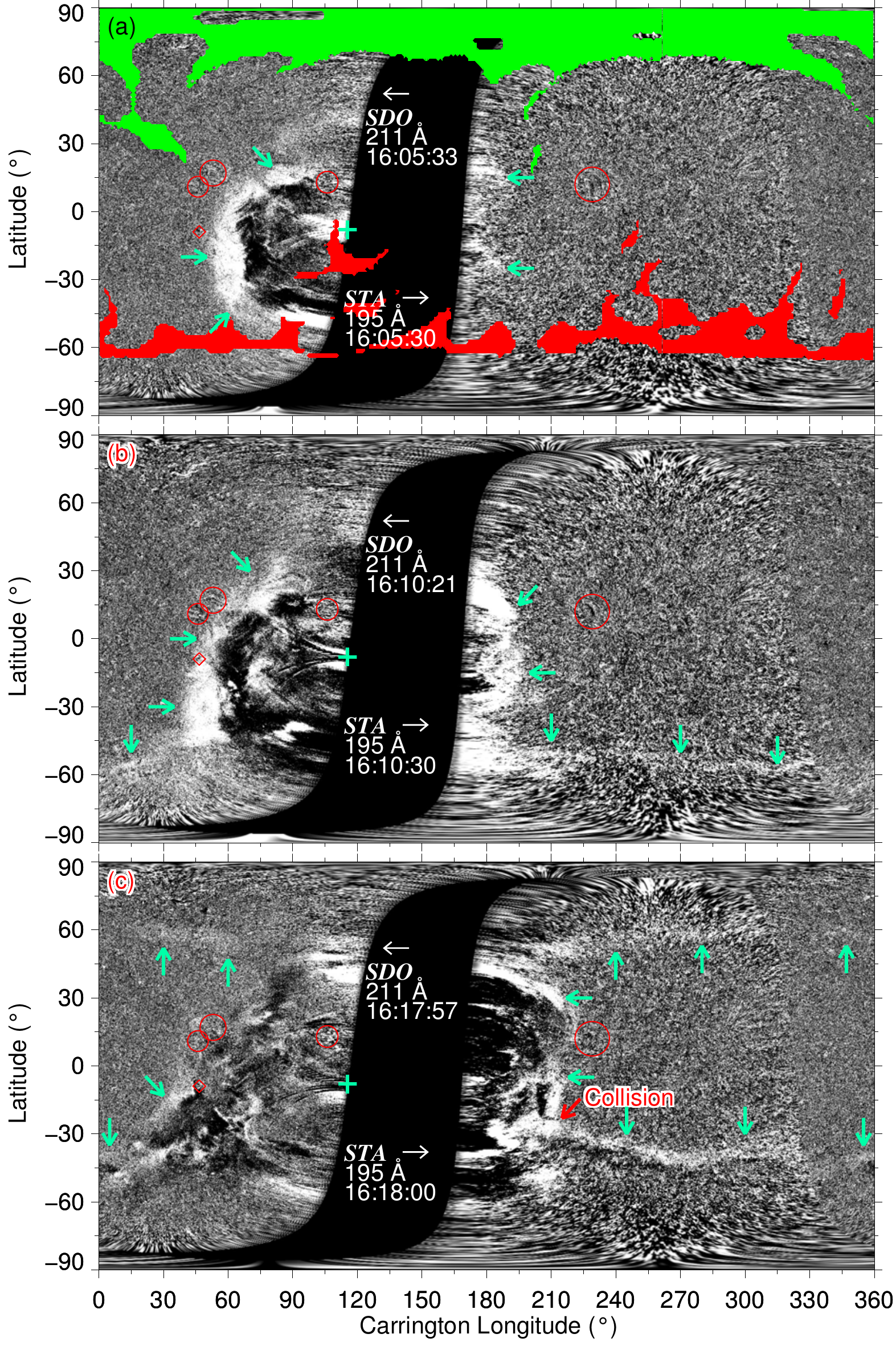}
\caption{\label{fig:synoptic}Synoptic maps constructed
from running-difference images of \sdo/AIA 211 \AA{} and \sta/EUVI 195 \AA.
The black region is unobservable from the two spacecraft.
The cyan arrows indicate the EUV wavefronts.
The cross symbol marks the position of the flare.
The diamond and circles represent the BP and ARs, respectively.
A GONG synoptic CH map is overlaid in panel (a),
where the green (red) regions denote the positive (negative) CHs.
The region below $\sim$$-$60\degr{} is uncovered by the GONG map
because the region is not well observed from the Earth.
The red arrow indicates a collision between the primary and transmitted EUV waves.
{In the animation of this figure,
the GONG synoptic CH map is indicated by green and red contours,
and the arrows are removed.
The animation begins at 15:33 UT and ends at 17:28 UT.}\\
(An animation of this figure is available.)}
\end{figure*}

Figure \ref{fig:synoptic} shows three synoptic maps
made by combining the RD images of AIA 211 \AA{} and EUVI-A 195 \AA{},
which illustrates the evolution of the EUV wave on the entire solar disk
{(despite the unobservable area)}
viewed from two opposite sides of the Sun.
Panel (a) shows that the primary EUV wave (indicated by the cyan arrows)
was observed by AIA and was just entering the FOV of EUVI-A around 16:05 UT.
The GONG synoptic plot marks the CHs covering the polar regions
above $\sim$60\degr{} north and south latitude.
The EUV wave is transmitted and also refracted by the CHs,
which changes the propagation direction to equatorward.
The south transmitted wave (indicated by the vertical arrows in Figure \ref{fig:synoptic}b)
emanates roughly along the boundary of the south-polar CH.
The transmitted waves are traveling antipoleward along all meridians from the polar CHs
and are observed simultaneously by \sdo{} and \sta,
as displayed in Figure \ref{fig:synoptic}(c).
The red arrow indicates a collision between the primary and transmitted waves,
which is also visible in the animation of Figure \ref{fig:synoptic}.
Corresponding parts of the primary wavefront are diffused when interacting with
the remote ARs (marked with circles in Figure \ref{fig:synoptic}),
and no obvious EUV wave is detected beyond the ARs
(except AR 12674 that is close to the flare site).
The primary EUV wave extends to a scale of $\sim$180\degr{}
($\sim$30\degr{}--$\sim$210\degr) in the longitudinal direction
before it becomes too faint to be detected.
The transmitted waves travel away from the polar CHs
and reach the equator on the opposite side of the flare.
The EUV wave is reflected and transmitted by both the north- and south-polar CHs,
and propagates through the 360\degr{} solar disk,
as simultaneously observed by \sdo{} and \sta{}
from two almost opposite sides of the Sun,
which is an unprecedented event.

\section{Conclusions and Discussions} \label{sec:concl}
We have investigated the sustained, global EUV wave
{and its connection with the accompanied CME-driven shock}
associated with the 2017 September 10 eruption,
using the observations from \sdo{}, \sta{}, {\goesr{} and \soho}.
The EUV wave is reflected and also transmitted by both the north- and south-polar CHs.
Two transmitted waves emanate respectively from the two polar CHs
and then propagate through the entire solar disk,
which is unprecedentedly observed from two nearly opposite sides of the Sun.
The reflection and transmission,
{which have also been observed by \cite{LiuJD2018ApJL} using single-spacecraft observations,}
are consistent with the fast-mode wave/shock interpretation for an EUV wave.
{We have examined the connection with the accompanied CME-driven shock
and the interactions with multiple coronal structures of the EUV wave.}
Below we conclude the results and discuss
the significant effects of coronal density and magnetic field distributions
on the globalization of the EUV wave
{and how the global EUV wave is connected with the CME-driven shock}.

\begin{enumerate}
    \item {After the CME-driven shock has propagated to the opposite side of the eruption,
    the shock is incurvated to the Sun and still connected with the CH transmitted EUV wave.
    The 3D ellipsoid model reveals that the primary EUV wave
    is the footprint of the CME-driven shock at the initial stage.
    After the shock has reached the opposite side,
    the incurvated shock forms a ``funnel-like'' structure and is still connected with the transmitted EUV wave
    (see Figure \ref{fig:backward}),
    although the northern part of the shock is not clearly observed by \sta/COR1.
    The primary EUV wave is still on the western hemisphere
    even after the shock has already propagated to the opposite side of the eruption.
    The transmitted EUV waves in the polar regions propagate much faster than the primary EUV wave
    in low- and midlatitudes (see the animation of Figure \ref{fig:euv_pfss}).
    If the primary EUV wave is also connected with the shock on the opposite side,
    the part of the shock coupled with the \emph{transmitted} wave may also propagate faster
    than its counterpart connected with the \emph{primary} wave.
    However, the connection between the shock and the primary EUV wave in our case
    cannot be identified by imaging observations from within the ecliptic plane,
    which may be determined by numerical simulations in the future.
    Additionally, a lateral collision between two streamers is observed
    (see {Figure \ref{fig:backward} and its animation}),
    which suggests that the lateral parts of the shock probably also collide on the opposite side.
    These results significantly improve the understanding
    of the connection between an EUV wave and the associated CME-driven shock
    on the opposite side of an eruption.}

    \item The transmitted EUV wave from the south-polar CH is accelerated
    inside an on-disk low-density dim region with closed magnetic fields,
    besides being accelerated inside the CH.
    {A plausible coronal cavity without a filament signature
    above the dim region was observed by \sdo/AIA
    when the dim region was traversing the east limb.}
    This kind of acceleration inside {an on-disk,} dim, closed field region
    has not been reported before to the best of our knowledge.
    In both the closed field region and the CH, the coronal density is relatively low
    and the local characteristic wave speed is high,
    although the magnetic field configurations are different.
    This implies that the acceleration of {an EUV wave}
    can be produced in both open and closed field regions
    as long as their density is low enough.

    \item Part of the primary EUV wavefront turns around a BP
    when the wavefront approaches {a filament channel} near the BP.
    The BP has a bipolar magnetic field structure similar to a small AR,
    {which halts part of the primary wave that interacts with the BP.
    The EUV wave, inside the low-density filament channel}
    with a high characteristic wave speed, is accelerated.
    The combination of {the halt and the acceleration
    caused by the BP and the filament channel}
    eventually changes the direction of the primary wavefront.
    {Speed elevation in a coronal cavity hosting a filament on the limb has been observed
    \citep{LiuON2012ApJ}.
    Our case demonstrates the effect of an on-disk filament channel
    and a small bipolar magnetic structure
    on the propagation of an EUV wave.}

    \item The primary EUV wavefront is diffused and apparently halted
    near the boundaries of remote ARs located 800 Mm away from the flare site,
    and no obvious AR transmitted secondary waves are observed beyond the ARs.
    \citet{ChenFC2016SoPh} indicate that a fast-mode wave could be converted to a slow-mode wave
    and then be trapped when it passes through a magnetic quasi-separatrix layer.
    \citet{OfmanT2002ApJ} demonstrate that transportation or dissipation
    of the wave energy can be a reason for a faint AR transmitted wave.
    The absence of obvious transmitted secondary waves in our case is different
    from previous cases with remote ARs within 400 Mm from the eruptions
    \citep[e.g.,][]{LiZY2012ApJ,ShenLS2013ApJ}.
    This suggests that the intensity of an AR transmitted secondary wave
    probably decreases with the distance {of the associated AR} from the eruption.
    {It is probably because the energy intensity of the EUV wave decreases
    with distance due to spread of the wave and/or energy dissipation.}

    \item {The EUV wave extends to an unprecedented scale of $\sim$360\degr{} in latitudes,
    which is mainly contributed by the polar CH transmission.
    The primary EUV wave persists for about 40 minutes and}
    extends to $\sim$180\degr{} in longitudes before it fades,
    even though it is probably restrained by the ARs on the eastern and western sides
    {(see Figure \ref{fig:synoptic})}.
    The CH transmitted EUV waves {appear $\sim$15 minutes later than the primary wave}
    and persist for over 50 minutes.
    The transmitted waves travel along all meridians and arrive at the opposite side of the eruption
    on the Sun
    (see {the animations of Figures \ref{fig:euv_pfss} and \ref{fig:synoptic}}),
    which is the major contributor to the unusually large extent and persistence of the EUV wave.
    If the EUV wave had not been transmitted by the two polar CHs,
    the wave would not have been so persistent and globalized to a 360\degr{} scale.

\end{enumerate}

\acknowledgments
{We are grateful to the anonymous referee for his/her valuable comments
that improved this paper.
We thank Dr. Xiaoshuai Zhu and Dr. Keiji Hayashi for their helpful discussions.}
The research is supported by the NSFC (grants 41774179, 41604146, and 41574140),
the Specialized Research Fund for State Key Laboratories of China,
and the CAS Strategic Priority Program on Space Science (XDA15011300).
H.H. is also supported by the China Scholarship Council (201804910106).
Z.Y. is also supported by the CAS Youth Innovation Promotion Association (2017188)
and the Beijing Natural Science Foundation (grant 1192018).
We acknowledge the use of data from \sdo{}, \stereo{}, \goesr{}, \soho, and the GONG program.

\end{CJK}
\end{document}